\documentclass[manuscript]{aastex}

\shorttitle{Ice-covered planets with high-pressure ice}
\shortauthors{Ueta \& Sasaki}

\begin{document}

\title{Structure of surface-H$_2$O layers of ice-covered planets with high-pressure ice}
\author{S. Ueta and T. Sasaki}
\affil{Earth and Planetary Sciences, Tokyo Institute of Technology, 2-12-1 Ookayama, Meguro-ku, Tokyo 152-8551, Japan}
\email{ueta@geo.titech.ac.jp, takanori@geo.titech.ac.jp}

\begin{abstract}
Many extrasolar (bound) terrestrial planets and free-floating (unbound) planets have been discovered. The existence of bound and unbound terrestrial planets with liquid water is an important question, and of particular importance is the question of their habitability. Even for a globally ice-covered planet, geothermal heat from the planetary interior may melt the interior ice, creating an internal ocean covered by an ice shell. In this paper, we discuss the conditions that terrestrial planets must satisfy for such an internal ocean to exist on the timescale of planetary evolution. The question is addressed in terms of planetary mass, distance from a central star, water abundance, and abundance of radiogenic heat sources. In addition, we investigate the structures of the surface-H$_2$O layers of ice-covered planets by considering the effects of ice under high pressure (high-pressure ice). As a fiducial case, $1M_{\oplus}$ planet at 1 AU from its central star and with 0.6 to 25 times the H$_2$O mass of Earth could have an internal ocean. We find that high-pressure ice layers may appear between the internal ocean and the rock portion on a planet with an H$_2$O mass over 25 times that of Earth. The planetary mass and abundance of surface water strongly restrict the conditions under which an extrasolar terrestrial planet may have an internal ocean with no high-pressure ice under the ocean. Such high-pressure-ice layers underlying the internal ocean are likely to affect the habitability of the planet.
\end{abstract}


\section{Introduction}

Since the first extrasolar planet was discovered in 1995 \citep{may95}, more than 800 exoplanets have been detected as of March 2013, owing to improvements in both observational instruments and the methods of analysis. Although most known exoplanets are gas giants, estimates based on both theory and observation indicates that terrestrial planets are also common \citep{how10}. Supporting these estimates is the fact that Earth-like planets have indeed been discovered. Moreover, space telescopes (e.g., {\it Kepler}) have now released observational data about many terrestrial-planet candidates. Whether terrestrial planets with liquid water exist is an important question to consider because it lays the groundwork for the consideration of habitability.

The orbital range around a star for which liquid water can exist on a planetary surface is called the habitable zone (HZ) (Hart 1979; Kasting et al. 1993). The inner edge of the HZ is determined by the runaway greenhouse limit \citep{kas88,nak92}, and the outer edge is estimated from the effect of CO$_2$ clouds \citep{kas93,mis00}. The region between these edges is generally called the HZ for terrestrial planets with plentiful liquid water on the surface (ocean planets). Planets with plentiful water on the surface but outside the outer edge of the HZ would be globally covered with ice, and no liquid water would exist on the surface. These are called "snowball planets" (Tajika 2008). Moreover, an ocean planet could be ice-covered even within the HZ because multiple climate modes are possible, including ice-free, partially ice-covered, and globally ice-covered states \citep{bud69,sel69,taj08}.  Although such planets would be globally ice-covered, liquid water could exist beneath the surface-ice shell if sufficient geothermal heat flows up from the planetary interior to melt the interior ice. In this scenario, only a few kilometers of ice would form at the surface of the ocean \citep{hof02}, and life could exist in the liquid water under the surface-ice shell \citep{hof98,hof02,gai99}.

Another possibility is presented by planets that float in space without being gravitationally bound to a star (free-floating planets), as have been found thanks to recent advances in observational techniques \citep{sum11}. Although such planets receive no energy from a central star, even a free-floating Earth-sized planet with considerable geothermal heat could have liquid water under an ice-covered surface.

Considering geothermal heat from the planetary interior, \citet{taj08} discusses the theoretical restrictions for ice-covered extrasolar terrestrial planets that, on the timescale of planetary evolution, have an internal ocean. \citet{taj08} shows that an internal ocean can exist if the water abundance and planetary mass are comparable to those of Earth.  A planet with a mass less than $0.4M_{\oplus}$ cannot maintain an internal ocean. For a planet with mass $\ge 4M_{\oplus}$, liquid water would be stable either on the planetary surface or under the ice, regardless of the luminosity of the central star and of the planetary orbit.  These are important conclusions and have important implications for habitable planets.

In this paper, we extend the analysis of Tajika (2008) and vary the parameter values such as abundance of radiogenic heat sources and H$_2$O abundance on the surface. Although \citet{taj08} assumed that the mass ratio of H$_2$O on the planetary surface is the same as that on Earth (0.023 wt\%), the origin of water on the Earth is not apparent \citep{gen08} so it is possible that extrasolar terrestrial planets have some order of H$_2$O abundance. We investigate this possibility by varying the H$_2$O abundance in our simulation, and also check whether ice appears under H$_2$O layers under high-pressure conditions (see Section 2.2). Therefore, in this work, we consider the effect of high-pressure ice under an internal ocean and discuss its implications for habitability (see Section 4.2). With these considerations, we discuss the conditions required for bound and unbound terrestrial planets to have an internal ocean on the timescale of planetary evolution (owing to geothermal heat flux from the planetary interior). Our discussion further considers various planetary masses, distances from the central star, water abundances, and the abundances of radiogenic heat sources. Finally, taking into account the effects of high-pressure ice, we investigate the structure of surface-H$_2$O layers of ice-covered planets.

\section{Method}
\subsection{Numerical model}

To calculate the mass-radius relationships for planets with masses in the range 0.1 $M_{\oplus}$-10 $M_{\oplus}$, we adjust the planetary parameters. We assume
\begin{equation}
\frac{R}{R_{\oplus}} = \left( \frac{M}{M_{\oplus}} \right)^{0.27}
\end{equation}
as per \citet{val06}, where $R$ is the planetary radius and $M$ is the planetary mass. The subscript $\oplus$ denotes values for Earth. The mantle thickness, core size, amount of H$_2$O, average density, and other planetary properties are scaled according to this equation.

The planetary surfaces are assumed to consist of frozen H$_2$O and to have no continental crust. We define the planetary radius as $R$ $=$ $dw$ $+$ $l$, where $dw$ is the H$_2$O thickness and $l$ is the mantle-core radius (see Fig. 1). The mass of H${_2}$O on the planetary surface is given by
\begin{equation}
M_{sw} = \frac{4}{3}\pi \rho_w[ (dw + l)^3 - l^3 ], 
\end{equation}
where $\rho_w$ is the density of H$_2$O. We vary $M_{sw}$ from $0.1 M_{sw0}$ to $100 M_{sw0}$, where $M_{sw0}$ $=$ $0.00023M$ with the prefactor being the H$_2$O abundance of Earth (0.023 wt.\%).

Assuming that the heat flux $q$ is transferred from the planetary interior through the surface ice shell by thermal conduction, the ice thickness $dh$ can be obtained as   
\begin{equation}
dh = k_i\frac{T_{ib} - T_s}{q},
\end{equation}
where $k_i$ is the thermal conductivity of ice, $T_{ib}$ is the temperature at the bottom of the ice, and $T_s$ is the temperature at the surface. We assume that the surface ice is hexagonal ice (ice Ih). Between 0.5 K and 273 K, the thermal conductivity of ice Ih is known \citep{dil65,kli75,var78}. For temperatures greater than $\sim$ 25 K, it is given by \citet{kli80} as
\begin{equation}
k_i = \frac{567 \mathrm{[Wm^{-1}]}}{T}.
\end{equation}
To estimate $T_{ib}$, we assume that the melting line of H$_2$O is a straight line connecting (0 bar, 273 K) to (2072 bar, 251 K) in the linear pressure-temperature phase diagram. The temperature $T_{ib}$ can be estimated using
\begin{eqnarray}
T_{ib} &=& 273 - \frac{22 \mathrm{[K]}}{2072 \mathrm{[bar]}}p_{ib} \nonumber \\ 
   &=& 273 - \frac{22 \mathrm{[K]}}{2072 \mathrm{[bar]}}dh\rho_w g \times 10^{-5},
\end{eqnarray}
where $p_{ib}$ (bar) is the pressure at the bottom of the ice and $g$ is the gravitational acceleration on Earth.

Considering energy balance on the planetary surface, the planetary surface temperature $T_s$ is
\begin{equation}
\frac{(1 - A)}{4d^2}\frac{L}{4\pi D_0^2} + q = \varepsilon \sigma T_s^4,
\end{equation}
where $A$ is the planetary albedo, $d$ is the distance from the central star in AU, $L$ is the luminosity of the central star, $D_0 = 1.5 \times 10^{11}$ m is a distance of 1 AU in meters, $\varepsilon$ is the emissivity of the planet, and $\sigma = 5.67 \times 10^{-8}$ Wm$^{-2}$K$^{-4}$ is the Stefan-Boltzmann constant. We assume $A = 0.62$ and $\varepsilon = 1.0$ (i.e., the planetary atmosphere contains no greenhouse gases, which yields an upper estimate of the ice thickness). The increase in luminosity due to the evolution of the central star as a main sequence star \citep{gou81} is considered using
\begin{equation}
L(t) = \left[1 + \frac{2}{5}\left(1 - \frac{t}{t_\odot}\right)\right]^{-1}L_\odot,
\end{equation}
where $t_\odot = 4.7 \times 10^9$ years, and $L_\odot = 3.827 \times 10^{26}$ W.

From these models, we can obtain the H$_2$O thickness $dw$ and the ice thickness $dh$. The condition for terrestrial planets to have an internal ocean is
\begin{equation}
dw > dh.
\end{equation}

To estimate the geothermal heat flux $q$ through planetary evolution, we investigate the thermal evolution of terrestrial planets by using a parameterized convection model (Tajika ＆ Matsui 1992; McGovern ＆ Schubert 1989; Franck ＆ Bounama 1995; von Blow et al. 2008; see appendix for details). We assume $E$, which is the initial heat generation per unit time and volume, is 0.1$E_0$ to 10$E_0$, where the constant $E_0$ is the initial heat generation estimated from the present heat flux of the Earth (see appendix for details).

\subsection{High-pressure ice}

Ice undergoes a phase transition at high pressure (Fig. 2). Unlike ice I, the other phases are more dense than liquid H$_2$O. We call the denser ice "high-pressure ice." Because \citet{taj08} assumes that the amount of H$_2$O on the planetary surface is the same as that on the Earth's surface $M_{sw0}$ ($= 0.00023M$), the only possible conditions on the planetary surface are those labeled 1, 2, and 3 in Fig. 3. However, because we consider herein that H$_2$O amounts may range from $0.1 M_{sw0}$ to $100 M_{sw0}$, the H$_2$O-rock boundary could move to higher pressure, so we should account for the effect of high-pressure ice (Fig. 3a). Therefore, types 4, 5, and 6 of Fig. 3b are added as possible surface conditions. Types-2 and type-5 planets both have an internal ocean, but high-pressure ice exists in type-5 planets between the internal ocean and the underlying rock.

We approximate the melting curve by straight lines connecting the triple points in the linear pressure-temperature phase diagram. We also assume that the amount of heat flux from the planetary interior that is transferred through the internal ocean by thermal convection is the same as that which is transferred through the surface ice. Here, we presume that temperature gradient in liquid-water part of phase diagram is isothermal (Figs. 3a and 3c), although a gradient for a deeper internal ocean than what is considered in this study should be carefully discussed. The condition for high-pressure ice to exist under the internal ocean is
\begin{equation}
P_h < P_b,
\end{equation}
where $P_b$ is the pressure at the H$_2$O-rock boundary and $P_h$ is the pressure on the phase diagram where the temperature gradient and the high-pressure melting line cross (Fig. 3c). As a representative value, we assume that high-pressure ice has a density of 1.2 g/cm$^3$. Because the characteristic features of high-pressure ice are poorly understood, we simplified the model; and, in particular, the thermal conductivities of high-pressure ice (see below). When thermal conductivity of high-pressure ice is relatively high and the temperature gradient in the high-pressure-ice part of the phase diagram is less than the gradient of the melting lines, the layer of high-pressure ice continues to the H$_2$O-rock boundary [(i) in Fig. 3c]. However, when the thermal conductivity is comparatively low and the temperature gradient is greater than that of the melting lines, the temperature gradient joins the melting line and goes along with the melting lines to the H$_2$O-rock boundary [dashed arrows (ii) and (iii) in Fig. 3c]. Although little is known about the conditions on the dashed arrows, we assume that layer to be high-pressure ice in this study. Therefore, from the point where the temperature-pressure line crosses into the high-pressure-ice part, the high-pressure layer continues to the H$_2$O-rock boundary.

\section{Results}

Figures 4a and 4b show the surface conditions for planets with masses from $0.1M_{\oplus}$ to $10M_{\oplus}$ at 4.6 billion years after planetary formation, with varying H$_2$O masses on the surface, with initial radiogenic heat sources, and at 1AU from our Sun. We assumed $E/E_0 = 1$ for Fig. 4a and $M_{sw}/M_{sw0} = 1$ for Fig. 4b. Because larger planets have larger geothermal heat flux and thicker H$_2$O layers, they could have an internal ocean with less H$_2$O mass on the planetary surface (Fig. 4a) and a weaker initial radiogenic heat source (Fig. 4b). However, larger planets also have larger gravitational acceleration. Thus, on those planets, high-pressure ice tends to appear under the internal ocean with smaller H$_2$O mass on the surface (Fig. 4a). For example, if a planet of mass of $1M_{\oplus}$ has an H$_{2}$O mass of $0.6M_{\oplus}$ to $25M_{sw0}$, it could have an internal ocean. However, if a planet has an H$_{2}$O mass $>25M_{sw0}$, high-pressure ice should exist under the ocean (Fig. 4a). Note, however, that an internal ocean can exist on a planet having a mass of $1M_{\oplus}$ if the initial radiogenic heat source exceeds $0.4E_0$ (Fig. 4b). Figures 4c and 4d give the temperature profiles of surface-H$_2$O layers. Figure 4c gives the conditions of a planet parameterized by $1M_{\oplus}$ and $5M_{sw0}$, whereas Fig. 4d is for $1M_{\oplus}$ and $30M_{sw0}$. Given the conditions of Fig. 4c, the surface H$_2$O layers consist of a conductive-ice-I layer and a convective-liquid layer (i.e., an internal ocean). When the planet has a greater H$_2$O mass, high-pressure ice could appear under the internal ocean (Fig. 4d). Here, the planetary surface temperature seems very low, so we assume that the planet is covered by ice that has a higher albedo ($\sim$0.6) than ocean/land ($\sim$0.3).

Figures 5a and 5b show the surface conditions for free-floating planets ($L = 0$) with masses from $0.1M_{\oplus}$ to $10M_{\oplus}$ at 4.6 billion years after planetary formation. The incident flux from the central star affects the surface temperature, thereby affecting the condition on the surface. Therefore, the conditions, and in particular those shown in Fig. 5a, are different from those shown in Figs. 4a and 4b. The results of Fig. 5a show that, regardless of the amount of H$_2$O a $1M_{\oplus}$ planet has, an internal ocean cannot exist under the ice shell. An internal ocean could exist on free-floating planets under certain conditions, but the planetary size and water abundance strongly constrain these conditions (see Fig. 5a). For instance, if a free-floating planet has an initial radiogenic heat source greater than $7E_0$, it can have an internal ocean (Fig. 5b). Figures 5c and 5d give the temperature gradient of surface-H$_2$O layers for free-floating planets. The parameters are set to the same values as those for Figs. 4c and 4b, except that $L = 0$. The temperature on the planetary surface, approximately 35 K, is calculated by Eq. (6) on the assumption that $L = 0$, and the ice-I layer is thicker than that for the conditions of Figs. 4c and 4d. Given the conditions of Fig. 5c, the surface-H$_2$O layers consist only of a conductive-ice-I layer. However, for grater H$_2$O mass, high-pressure ice could appear under the ice-I layer (Fig. 5d).

Figure 6 shows the surface conditions for planets with masses from $0.1M_{\oplus}$ to $10M_{\oplus}$ at varying distances from a central star, and at 4.6 billion years after planetary formation. The runaway greenhouse limit \citep{kas88,nak92} indicating the inner edge of the HZ is not considered.  The effect of the incident flux from the central star on the surface conditions is estimated in each graph of Fig. 6 as a function of distance from the central star. We find that the existence of an internal ocean on planets far from a central star depends on the planetary mass and surface-H$_2$O mass. For the conditions of Fig. 6b on which a planet has two times the H$_2$O mass (i.e., 2$M_{sw}/M_{sw0}$), for example, a $2M_{\oplus}$ planet can have an internal ocean under which there is no high-pressure ice only out to approximately 7 AU from the central star. For a planet with five times the H$_2$O mass (i.e., 5$M_{sw}/M_{sw0}$; Fig. 6c), an internal ocean with no underlying high-pressure ice can exist out to approximately 30 AU. However, if the planet has ten times the H$_2$O mass (i.e., 10$M_{sw}/M_{sw0}$; Fig. 6d), an internal ocean without underlying high-pressure ice could exist to only approximately 5 AU. The planetary mass and surface-H$_2$O mass strongly constrain the conditions under which an extrasolar terrestrial planet far from its central star can have an internal ocean with no underlying high-pressure ice.

\section{Discussion}

\subsection{Models of study}

We expanded on the models of Tajika (2008) by (1) invoking the mass-radius relationship [Eq. (1)] to consider planetary compression by gravity, (2) considering the temperature dependence of the thermal conductivity of ice Ih [Eq. (4)], and varying the (3) abundance of radiogenic heat sources and (4) H$_2$O abundance on the surface, both of which were held constant in Tajika (2008). We used Eq. (1) to determine the mass-radius relationships because we expected more accurate results, although this change makes little quantitative difference in the results. Below, we discuss items (2), (3), and (4) to analyze the models of this study.

It is known from experiment that the thermal conductivity of ice Ih depends on temperature [$\approx 1/T$ \citep{kli80}]. In the present study, the surface-ice shell is thicker than that considered by Tajika (2008). For example, the surface-ice shell considered herein is about 1.1 times thicker at 1 AU and about 3.5 to 4.4 times thicker on the free-floating planets ($L = 0$). This increased thickness is due to the use of Eq. (4) to describe the thermal conductivity of ice Ih in contrast with Tajika (2008), where a constant thermal conductivity ($k_{i} = 2.2$) was used. Tajika (2008) treated the abundance of H$_2$O and heat source elements (a ratio of mass against a planetary mass) to be constant, i.e., $M_{sw}/M_{sw0} = 1$ ($M_{sw0}$ $=$ $0.00023M$) and $E/E_0 = 1$, although the amounts should vary with a planetary mass, and showed that liquid water would be stable either on the surface or beneath ice for a planet with a mass exceeding $4M_{\oplus}$, regardless of planetary orbit and luminosity of the central star. In this paper, however, when we consider a planet with parameters $M_{sw}/M_{sw0} = 1$ and $E/E_0 = 1$, the results shown in Fig. 5 are not consistent with those of Tajika (2008). Our results indicate that free-floating planets with masses between $0.1M_{\oplus}$ and $10M_{\oplus}$ could not have liquid water on the planetary surface and also under the ice for planetary parameters of $M_{sw}/M_{sw0} = 1$ and $E/E_0 = 1$. As just explained, when using Eq. (4) to describe the thermal conductivity of ice Ih, the planetary surface temperature becomes more sensitive to the thickness of the surface-ice layer, which leads to a several-times increase in the thickness of the surface ice. These results thus differ from those of Tajika (2008).

In this study, we assume that the high-pressure layer continues to the H$_2$O-rock boundary from the point where the temperature-pressure line crosses into the high-pressure-ice part (see Section 2.2). The results of this paper could depend on this assumption. When the thermal conductivity of high-pressure ice is comparatively low (similar to that of ice Ih), the heat cannot be transported through the high-pressure ice effectively, thus the bottom of the high-pressure ice could be thermally unstable and be melted enough to form internal ocean.

The result of the present study indicate that the high-pressure layers under the internal ocean could be from $< \sim$1 km to $\sim$100 km thick because a super-Earth planet with a few wt.\% H$_2$O could have $\sim$100 km-thick high-pressure ice layers. In high-pressure ice that is $\sim$100 km thick, it is possible that convective ice layers appear. To estimate whether or not such convection layers arise, we use the Rayleigh number $R_a$, which is given by
\begin{equation}
R_a = \frac{g\alpha \rho a^3 \Delta T}{\kappa \eta},
\end{equation} 
where $\alpha$ is the coefficient of thermal expansion, $\rho$ is the density, $a$ is the thickness of the high-pressure-ice layer, $\Delta T$ is the temperature difference between top and bottom part of the layer, $\kappa$ is the thermal diffusivity, and $\eta$ is the viscosity coefficient. When the Rayleigh number $R_a$ exceeds the critical value for the onset of convection ($R_{a_{crit}}$ $\sim 10^3$), convective motion spontaneously begins. If we use the parameters from Kubo (2008), i.e., $\alpha = 10^{-4}$ K$^{-1}$, $\rho = 1000 $ kgm$^{-3}$, $\Delta T = 50$ K, $\kappa = 2 \times 10^{-6}$ m$^{2}$s$^{-1}$, and $\eta = 10^{15}-10^{18}$ Pa s, and use the typical values of this study, $g = 10$ ms$^{-2}$ and $a = 10^2$ km, the Rayleigh number is $\sim 10^4-10^7$, which indicates that the convective motion is possible.

In this study, we consider $M_{sw}$ from 0.1$M_{sw0}$ (0.0023 wt.\%) to 100$M_{sw0}$ (2.3 wt.\%). If the planet has significantly more H$_2$O on the surface, the water layer would be very thick and our model would not apply to that planet.  If the planet has significantly less H$_2$O on the surface, the regassing flux of water would change because of insufficient water on the surface, and the planet's thermal evolution would differ from that of Earth.  In other words, we consider only those planets that are relatively active geothermally and have an adequate amount of water as Earth-like planets.  Improving our model so that it applies to other types of terrestrial planets is an important problem that we leave for future work.

\subsection{Habitability of internal ocean}

For genesis and sustenance of life, we need at least (1) liquid water and (2) nutrient salts because these substances are required to synthesize the body of life (Maruyama et al. 2013). Because nutrient salts are supplied from rocks, it is necessary that liquid water should be in contact with rock to liberate the salts. A type-5 planet (Fig. 3b) is thus not likely to be habitable because the internal ocean does not come in contract with rocks. However, it is possible for a type-2 planet to meet this requirement. Therefore, we presume that only type-2 planets have an internal ocean that is possibly habitable.

Therefore, the results of this study indicate that planetary mass and H$_2$O mass constitute two more conditions to add to the previous conditions for an extrasolar planet to have an internal ocean without high-pressure ice. In other words, these considerations indicate that only a planet with the appropriate planetary mass and H$_2$O mass can have an internal ocean that is possibly habitable.

However, it is possible that hydrothermal activities within the rocky crust may transfer nutrient salts to the internal ocean through cracks in the high-pressure ice. Large terrestrial planets such as those considered in this study are likely to have areas of high geothermal activity along mid-oceanic ridges and subduction zones as well as large submarine volcanos whose tops might emerge into the internal ocean from the high-pressure ice. Furthermore, for planets with thick high-pressure-ice layers (Section 4.1), the convective ice layer could transfer nutrient salts from the rock to the internal ocean. In these cases, high-pressure ice might not prevent nutrient flux from the rock floor from reaching the internal ocean.

\subsection{Future work}

As shown in Figs. 4b and 5b, an appropriate initial radiogenic heat source is an important factor in determining whether or not a planet has an internal ocean. Because the variation in the amount of initial heat generation in planets throughout space is not known, we assume in this study that the initial heat generation $E$ per unit time and volume ranges from $0.1E_0$ to $10E_0$. Thus, in order to resolve this issue, the general amount of radiogenic heat sources for extrasolar terrestrial planets should be estimated.

An Earth-size planet ($M_{sw}/M_{sw0} = 1$, $E/E_0 = 1$) orbiting at 1AU around the Sun for 4.6 billion years with no greenhouse gases might be globally covered by ice (Fig. 4). In contrast, Earth currently is partially covered with ice but retains liquid water on its surface. Therefore, it is almost certain that greenhouse gases play a major role in keeping the surface warm. Even free-floating, Earth-sized planets with atmospheres rich in molecular hydrogen could have liquid water on the surface because geothermal heat from the interior would be retained by the greenhouse-gas effect of H$_2$ \citep{ste99}. By applying this model here, we can account for the greenhouse-gas effect for various values of emissivity $\varepsilon$ in Eq.(6). In future presentations, we will thus discuss how greenhouse gases modify the conditions necessary for the development of a terrestrial ocean planet or ice-covered planet with an internal ocean.

In the present study, we assumed pure H$_2$O on the planetary surface. However, even if an extrasolar terrestrial planet has surface H$_2$O, the H$_2$O might not be pure because it might contain dissolved nutrient salts such as those found in Earth's oceans. In this case, the melting point of the solution would differ from that of pure H$_2$O. Moreover, the phase diagram would become more complicated, and the properties of H$_2$O (e.g., thermal conductivity) could be transmuted as a result of the appearance of phases in which H$_2$O ice contains nutrient salts. A very important work would thus be to analyze multicomponent water (e.g., seawater on Earth) to see what qualitative changes such a modified phase diagram would bring to our results.

Consider the examples, Europa and Ganymede, which are the two satellites of Jupiter and are thought to have internal oceans. Ganymede is thought to have high-pressure ice under its internal ocean (see, e.g., Lupo 1982). Therefore, determining whether the internal oceans of Europa and Ganymede are suitable for life would be pertinent to the discussion of the habitability of internal oceans with or without high-pressure ice.  In addition, a more circumstantial discussion of the habitability of internal oceans would require considering the redox gradient within the internal ocean [as exemplified by the discussions of Europa (Gaidos et al. 1999)], and the effects of the riverine flux of nutrient salts (Maruyama et al. 2013).

\section{Conclusion}

Herein, we discuss the conditions that must be satisfied for ice-covered bound and unbound terrestrial planets to have  an internal ocean on the timescale of planetary evolution. Geothermal heat flow from the planetary interior is considered as the heat source at the origin of the internal ocean. By applying and improving the model of \citet{taj08}, we also examine how the amount of radiogenic heat and H$_2$O mass affect these conditions. Moreover, we investigate the structures of surface-H$_2$O layers of snowball planets by considering the effects of high-pressure ice. The results indicate that planetary mass and surface-H$_2$O mass strongly constrain the conditions under which an extrasolar terrestrial planet might have an internal ocean without a high-pressure ice existing under the internal ocean.\\

\acknowledgments
We thank the reviewer Eiichi Tajika for the constructive comments.  We also thank S. Ida and M. Ikoma for valuable discussions.  This research was supported by a grant for the Global COE Program "From the Earth to 'Earths'" from the Ministry of Education, Culture, Sports, Science and Technology of Japan.  T. S. was supported by a Grant-in-Aid for Young Scientists (B), JSPS KAKENHI Grant Number 24740120.

\appendix
\section{Parameterized convection model}
By applying conservation of energy, we obtain
\begin{equation}
4\pi R_m^2q + \frac{4}{3}\pi \rho c(R_m^3 - R_c^3)\frac{dT_m}{dt} = \frac{4}{3}\pi E(t)(R_m^3 - R_c^3),
\end{equation}
where $\rho$ is the density of the mantle, $c$ is the specific heat at constant pressure, $R_m$ and $R_c$ are the outer and inner radii of the mantle, respectively, $T_m$ is the average mantle temperature, and $E(t)$ is the rate of energy production by decay of radiogenic heat sources in the mantle per unit volume.

The mantle heat flux is parameterized in terms of the Rayleigh number $R_a$ as
\begin{equation}
q = \frac{k(T_m - T')}{R_m - R_c}\left(\frac{R_a}{R_{a_{crit}}}\right)^\beta,
\end{equation}
where $k$ is the thermal conductivity of the mantle, $T'$ is the temperature at the surface of the mantle, $R_{a_{crit}}$ is the critical value of $R_a$ for convection onset, and $\beta$ is an empirical constant.

The radiogenic heat source is parameterized as
\begin{equation}
E(t) = Ee^{-\lambda t}
\end{equation}
where $\lambda$ is the decay constant of the radiogenic heat source and $E$ is the initial heat generation per unit time and volume. We assume $E$ is 0.1$E_0$ to 10$E_0$, where the constant $E_0$ is the initial heat generation estimated from the present heat flux of Earth.

We obtain $R_a$ as
\begin{equation}
R_a = \frac{g\alpha(T_m - T')(R_m - R_c)^3}{\kappa \nu},
\end{equation} 
where $\alpha$ is the coefficient of thermal expansion, $\kappa$ is the thermal diffusivity, and $\nu$ is the water-dependent kinematic viscosity. The viscosity $\nu$ strongly depends on the evolution of the mass $M_w$ of mantle water and the mantle temperature $T_m$ and is parameterized as
\begin{equation}
\nu = \nu_m\exp(T_A/T_m),
\end{equation}
\begin{equation}
T_A = \alpha_1 + \alpha_2x,
\end{equation}
\begin{equation}
x = \frac{M_w}{M_m},
\end{equation}
where $\nu_m$, $\alpha_1$, and $\alpha_2$ are constants, $T_A$ is the activation temperature for solid-state creep, and $M_m$ is the mass of the mantle.

The evolution of the mantle water can be described by the regassing flux $F_{reg}$ and outgassing flux $F_{out}$ as
\begin{eqnarray}
\frac{dM_w}{dt} &=& F_{reg} - F_{out} \nonumber \\ 
                     &=& f_{bas}\rho_{bas}d_{bas}SR_{H_2O} - \frac{M_w}{\frac{4}{3}\pi (R_m^3 - R_c^3)}d_mf_wS,
\end{eqnarray}
where $f_{bas}$ is the water content in the basalt layer, $\rho_{bas}$ is the average density, $d_{bas}$ is the average thickness of the basalt layer before subduction, $S$ is the areal spreading rate, $R_{H_2O}$ is the regassing ratio of water, $d_m$ is the melting generation depth, and $f_w$ is the outgassing fraction of water. The regassing ratio of water linearly depends on the mean mantle temperature $T_m$ via
\begin{equation}
R_{H_2O} = R_T(T_m(0) - T_m) + R_{H_2O,0}
\end{equation}
as given by von Bloh et al. (2008), where $R_T$ is the temperature dependence of the regassing ratio, and $R_{H_2O,0}$ is the initial regassing ratio. The areal spreading rate $S$ is
\begin{equation}
S = \frac{q^2\pi \kappa A_0}{4k^2(T_m - T')^2},
\end{equation}
where $A_0$ is the ocean-basin area. The area $A_0$ can be parameterized as
\begin{equation}
\frac{A_0}{A_0^*} = \left( \frac{R}{R_{\oplus}} \right)^2,
\end{equation}
where $A_0^*$ is the present ocean-basin area on Earth. By using Eq. (1), we obtain $A_0$ as
\begin{equation}
A_0 = A_0^*\left( \frac{M}{M_{\oplus}} \right)^{0.54}.
\end{equation}

In Table 1, we summarize selected values for the parameters used in the parameterized convection model.

\clearpage

\begin{figure}
\centering
\includegraphics[width=10cm,bb= 0 0 960 720,trim=0 0 0 180]{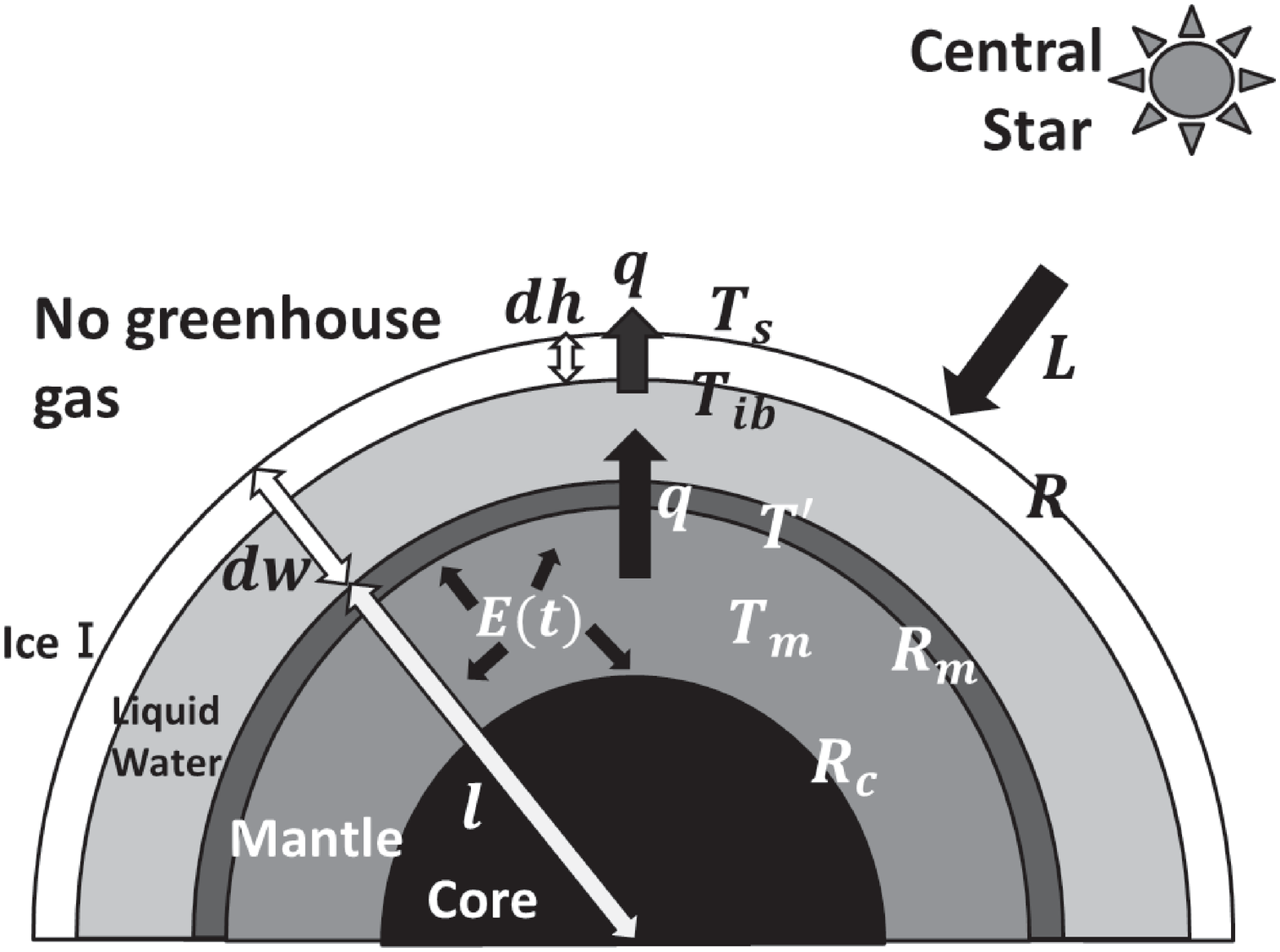}
\caption{Schematic of model used in this study}
\end{figure}

\clearpage

\begin{figure}
\centering
\includegraphics[width=10cm,bb= 0 0 960 720,trim=0 0 0 180]{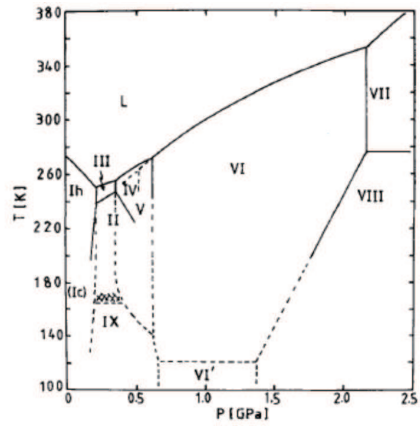}
\caption{Phase diagram of H$_2$O [after \citet{kam73}]}
\end{figure}

\clearpage

\begin{figure}
\centering
\includegraphics[width=10cm,bb= 0 0 960 720,trim=0 0 0 180]{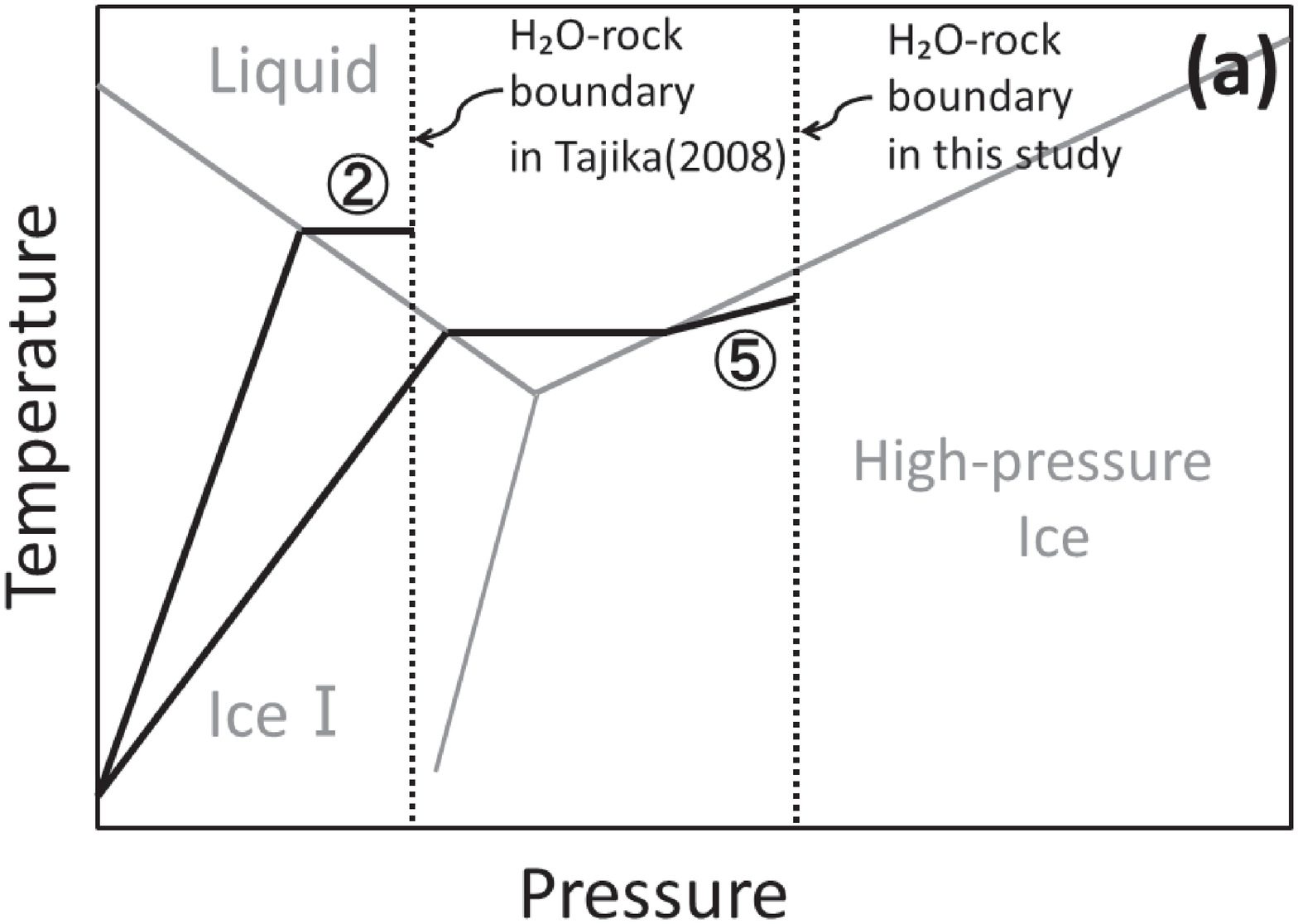}
\includegraphics[width=10cm,bb= 0 0 960 720,trim=0 0 0 180]{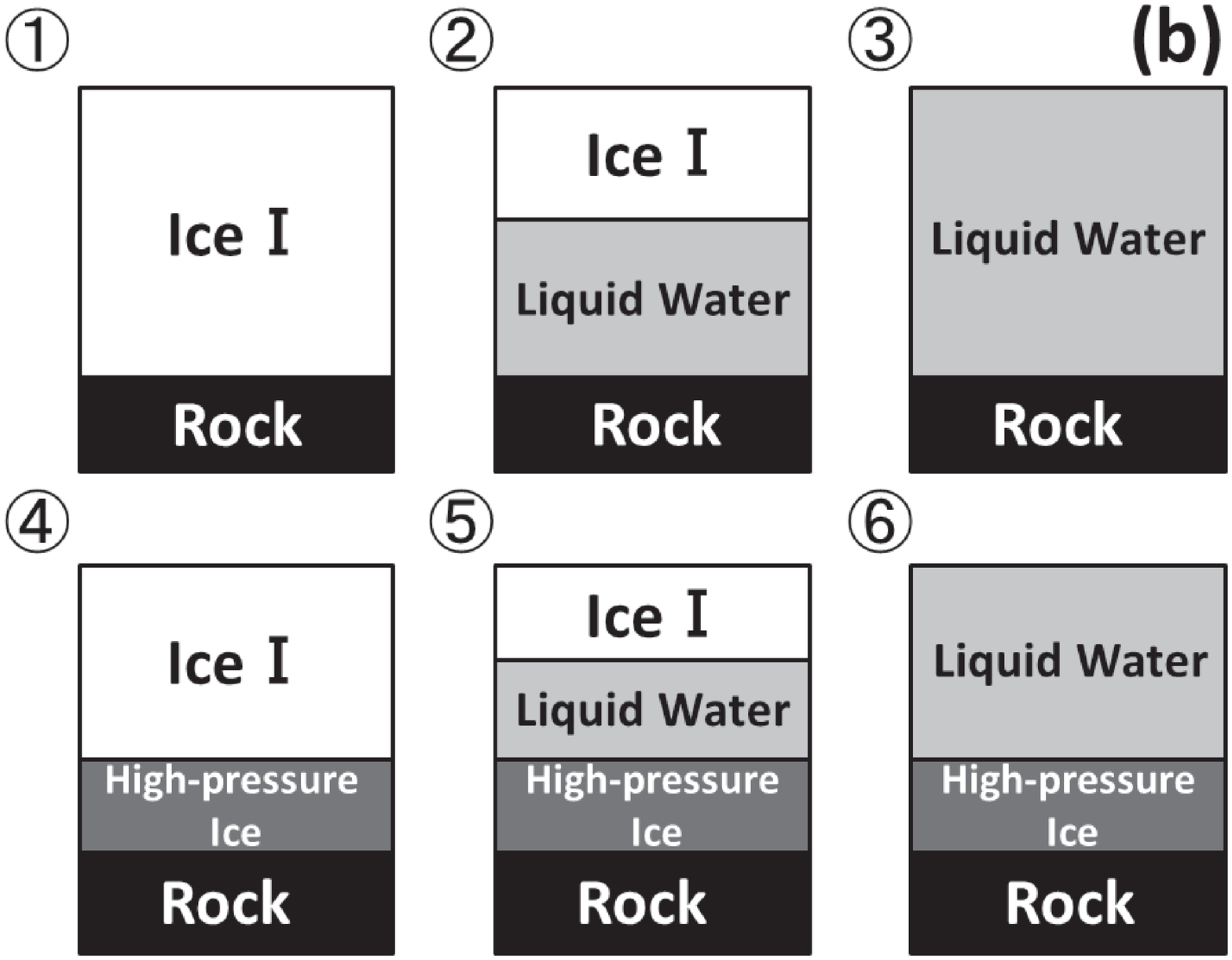}
\includegraphics[width=10cm,bb= 0 0 960 720,trim=0 0 0 180]{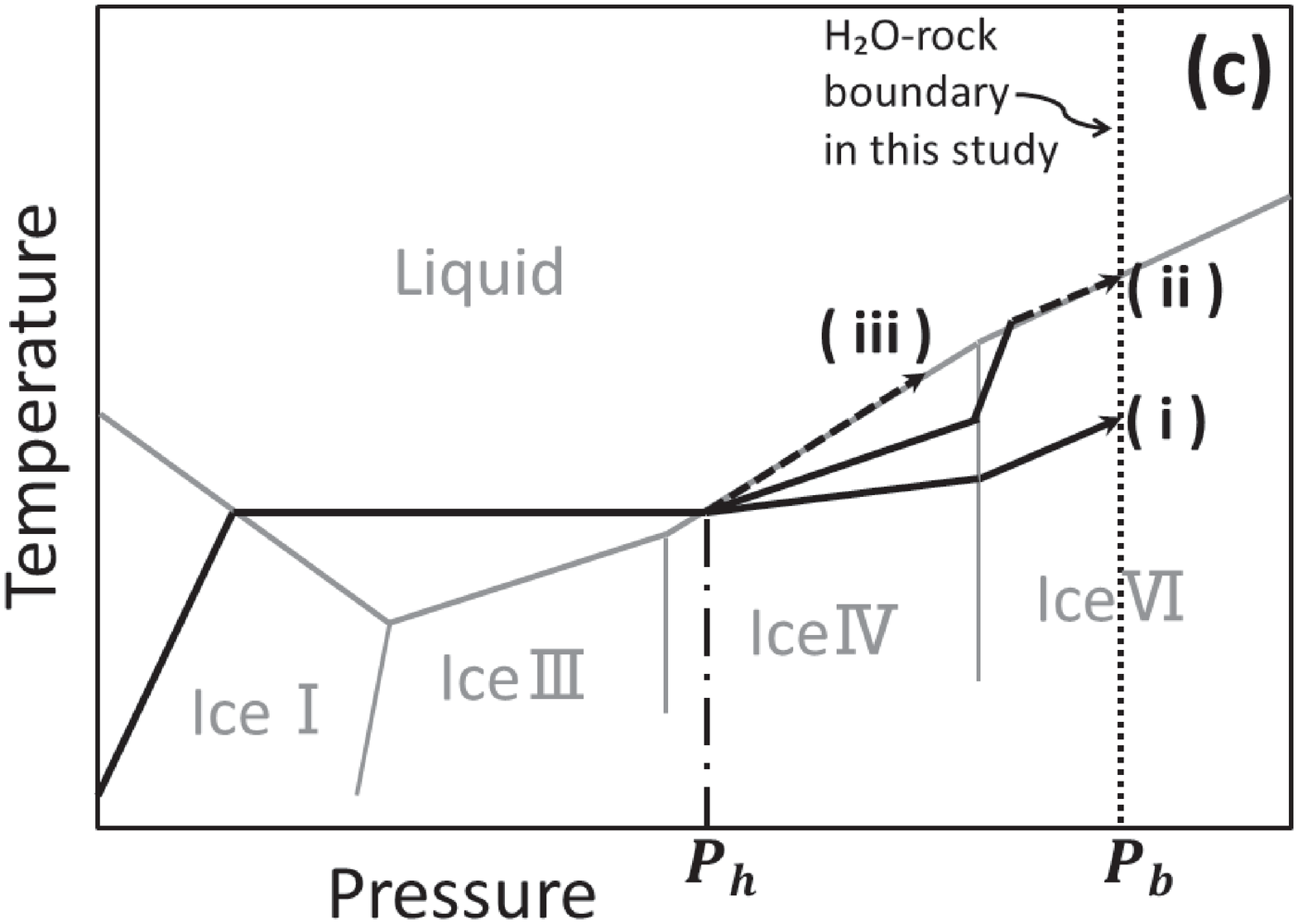}
\caption{(a) Schematic phase diagram of H$_2$O (gray lines), temperature gradient (black lines), and H$_2$O-rock boundaries (dashed lines). (b) Types of planets that have H$_2$O on the planetary surface. (c) Schematic of models used to treat high-pressure ice.}
\end{figure}

\clearpage

\begin{figure}
\centering
\includegraphics[width=8cm,bb= 0 0 960 720,trim=0 0 0 180]{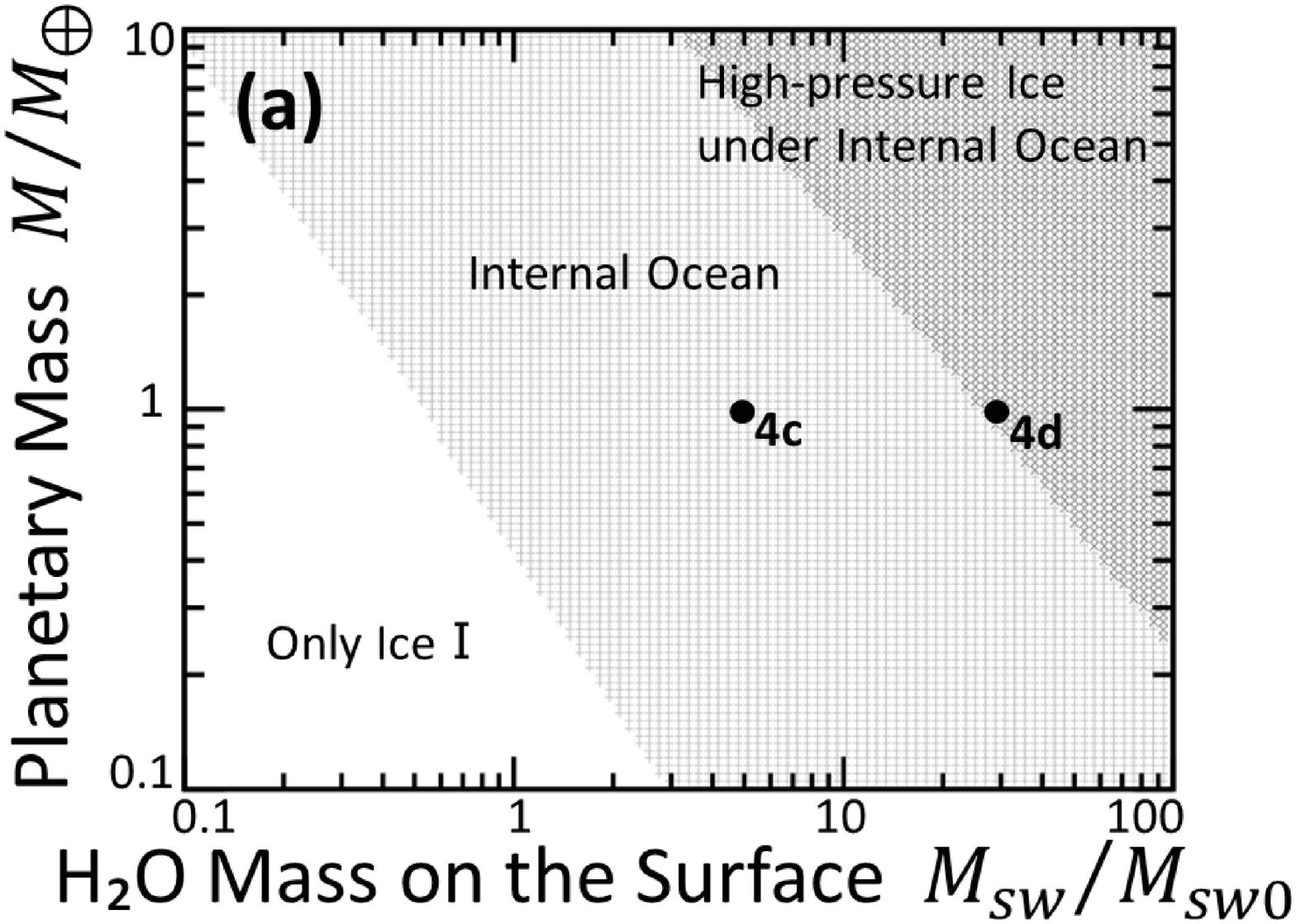}
\includegraphics[width=8cm,bb= 0 0 960 720,trim=0 0 0 180]{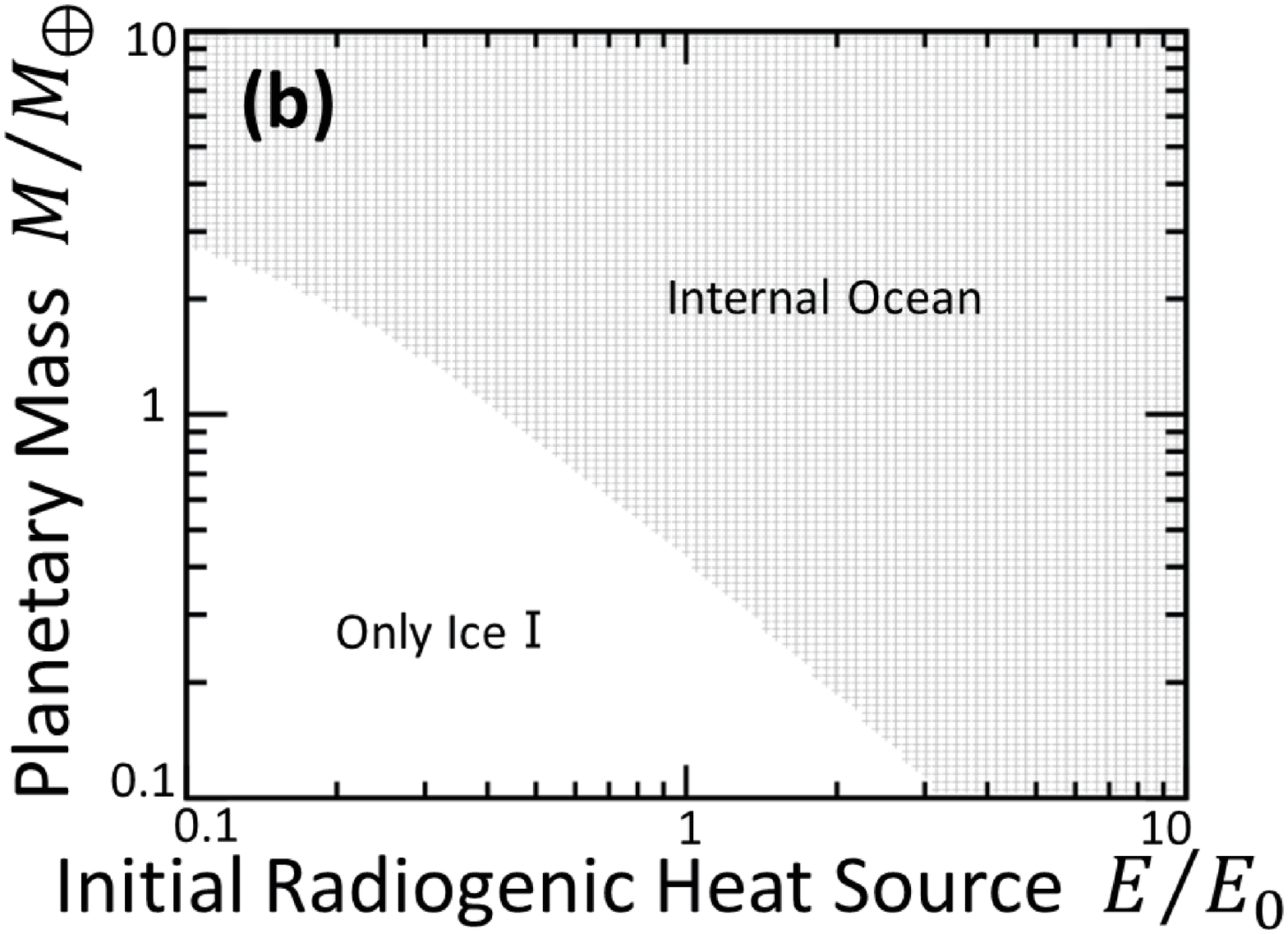}
\includegraphics[width=8cm,bb= 0 0 960 720,trim=0 0 0 180]{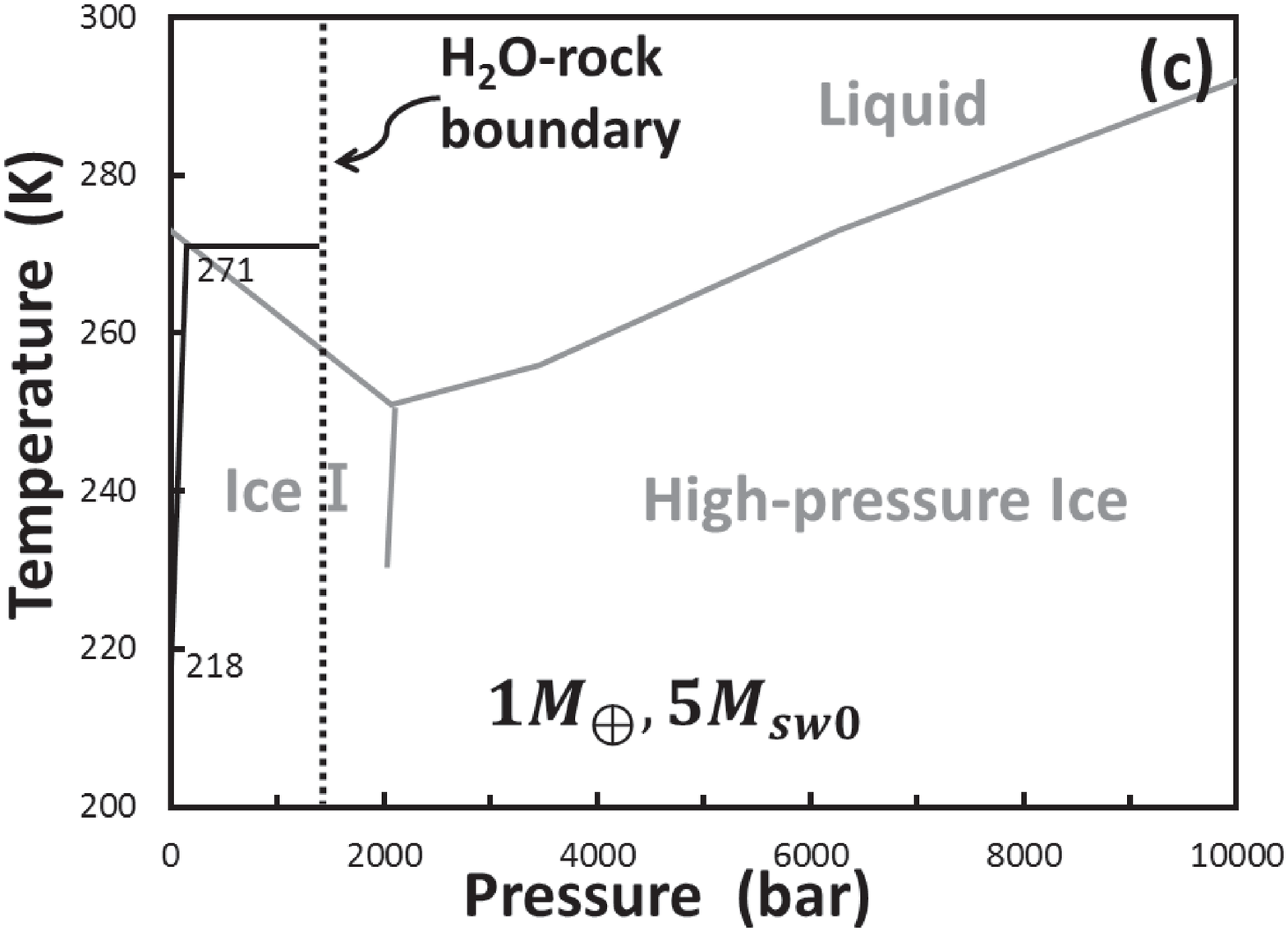}
\includegraphics[width=8cm,bb= 0 0 960 720,trim=0 0 0 180]{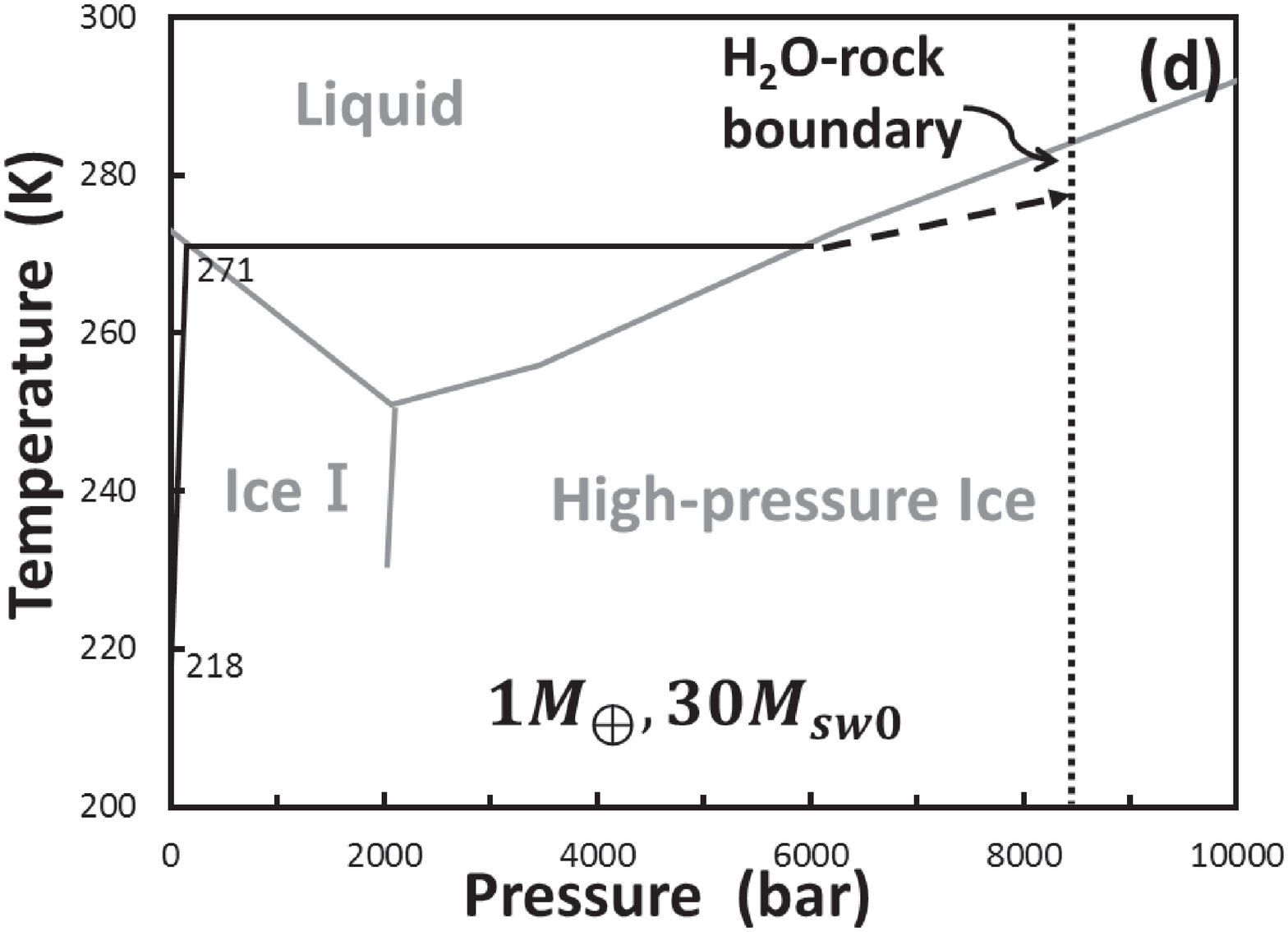}
\caption{Surface conditions for planet at 1AU around a central star ($L=L_0$; the present luminosity of our Sun) and temperature profiles of surface-H$_2$O layers for two cases. (a) Horizontal axis is surface-H$_2$O mass, and vertical axis is planetary mass normalized by Earth's mass, assuming $E/E_0 = 1$. The two dots in this panel indicate the conditions corresponding to the temperature profiles shown in panels (c) and (d). (b) Horizontal axis is initial radiogenic heat, and vertical axis is the planetary mass normalized by Earth's mass, assuming $M_{sw}/M_{sw0} = 1$. (c) Temperature profile at $1M_{\oplus}$ and $5M_{sw0}$. (d) Temperature profile at $1M_{\oplus}$ and $30M_{sw0}$.}
\end{figure}

\clearpage

\begin{figure}
\centering
\includegraphics[width=8cm,bb= 0 0 960 720,trim=0 0 0 180]{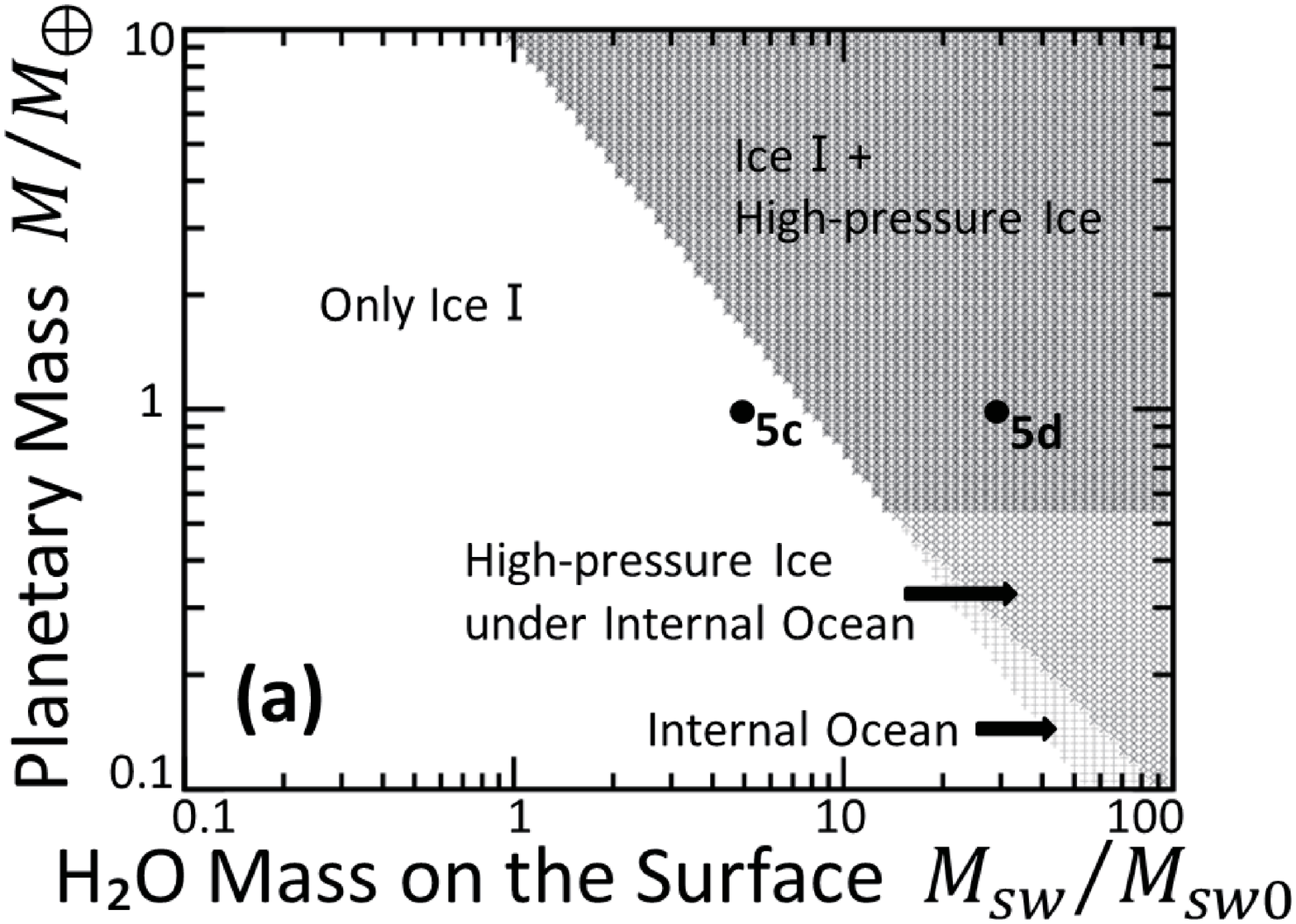}
\includegraphics[width=8cm,bb= 0 0 960 720,trim=0 0 0 180]{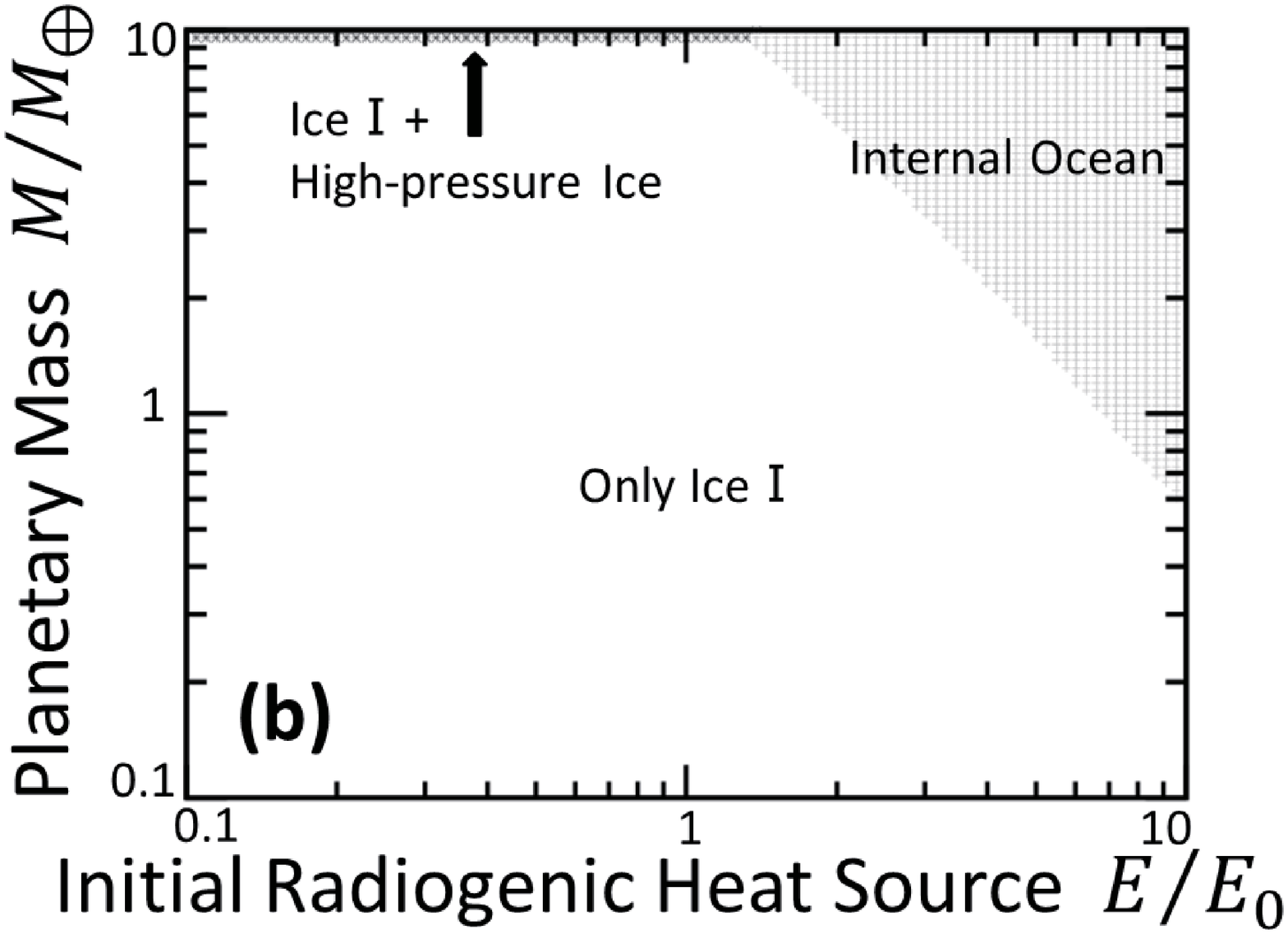}
\includegraphics[width=8cm,bb= 0 0 960 720,trim=0 0 0 180]{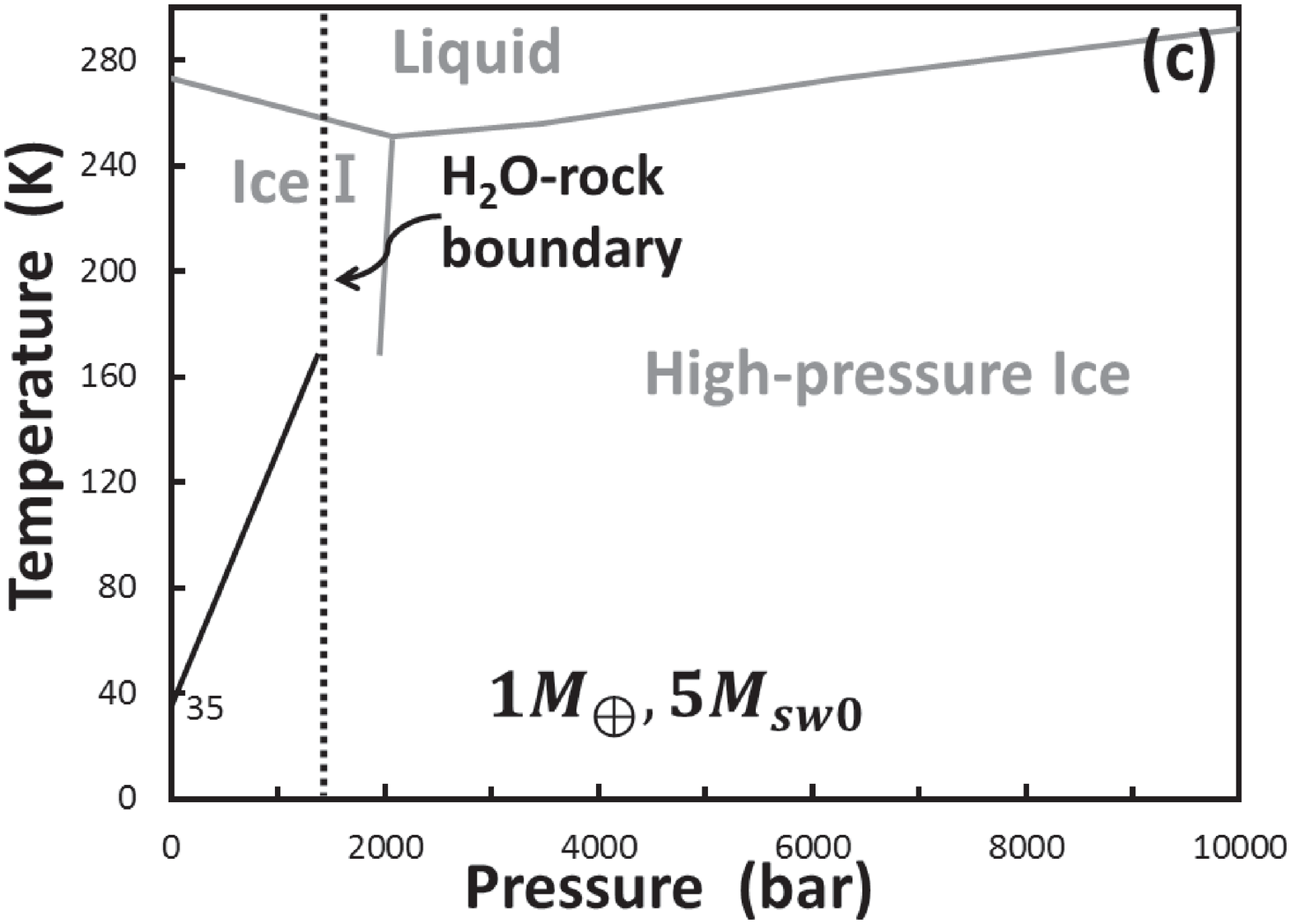}
\includegraphics[width=8cm,bb= 0 0 960 720,trim=0 0 0 180]{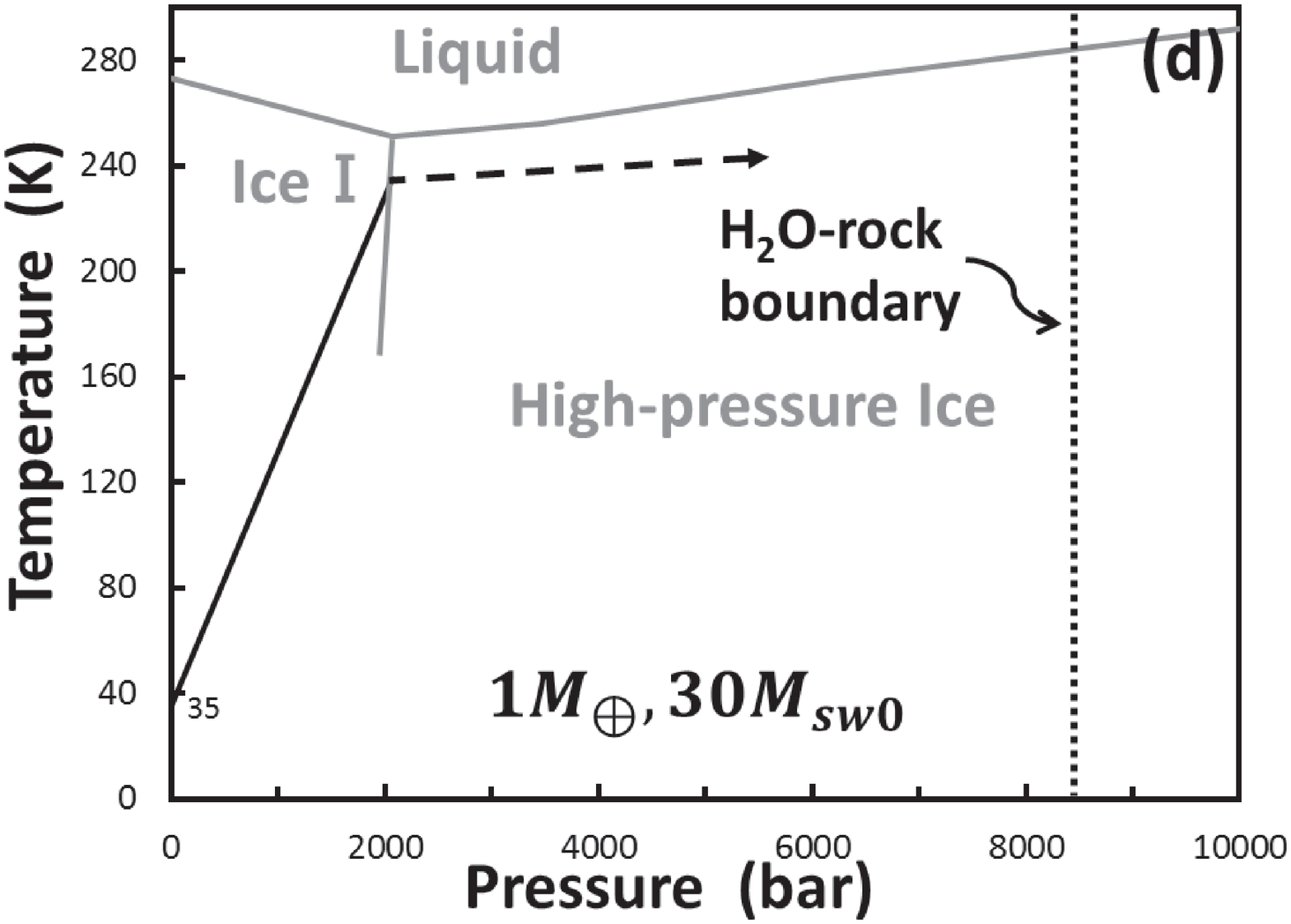}
\caption{Same as Fig. 4, but for a flee-floating planet ($L = 0$).}
\end{figure}

\clearpage

\begin{figure}
\centering
\includegraphics[width=8cm,bb= 0 0 960 720,trim=0 0 0 180]{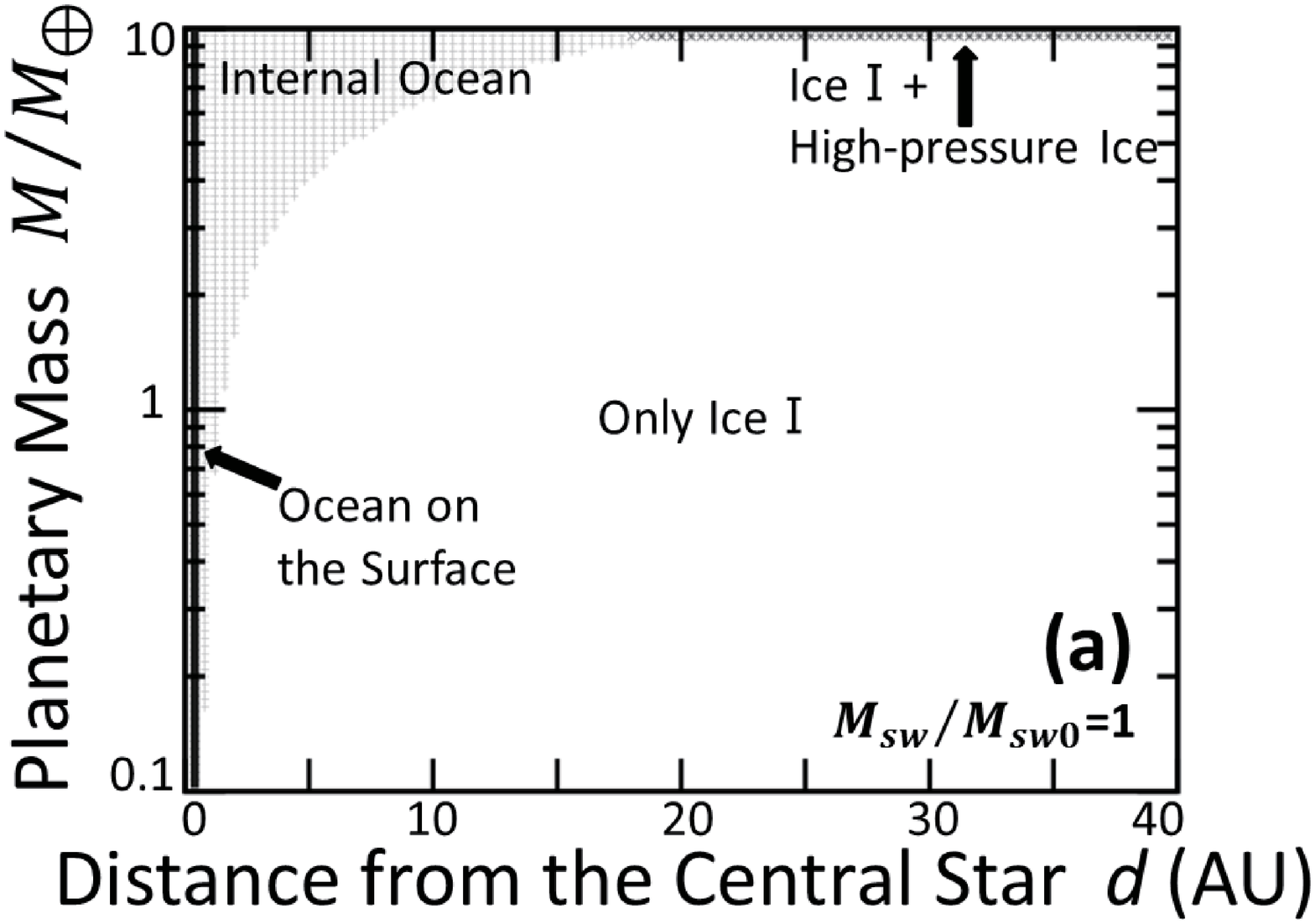}
\includegraphics[width=8cm,bb= 0 0 960 720,trim=0 0 0 180]{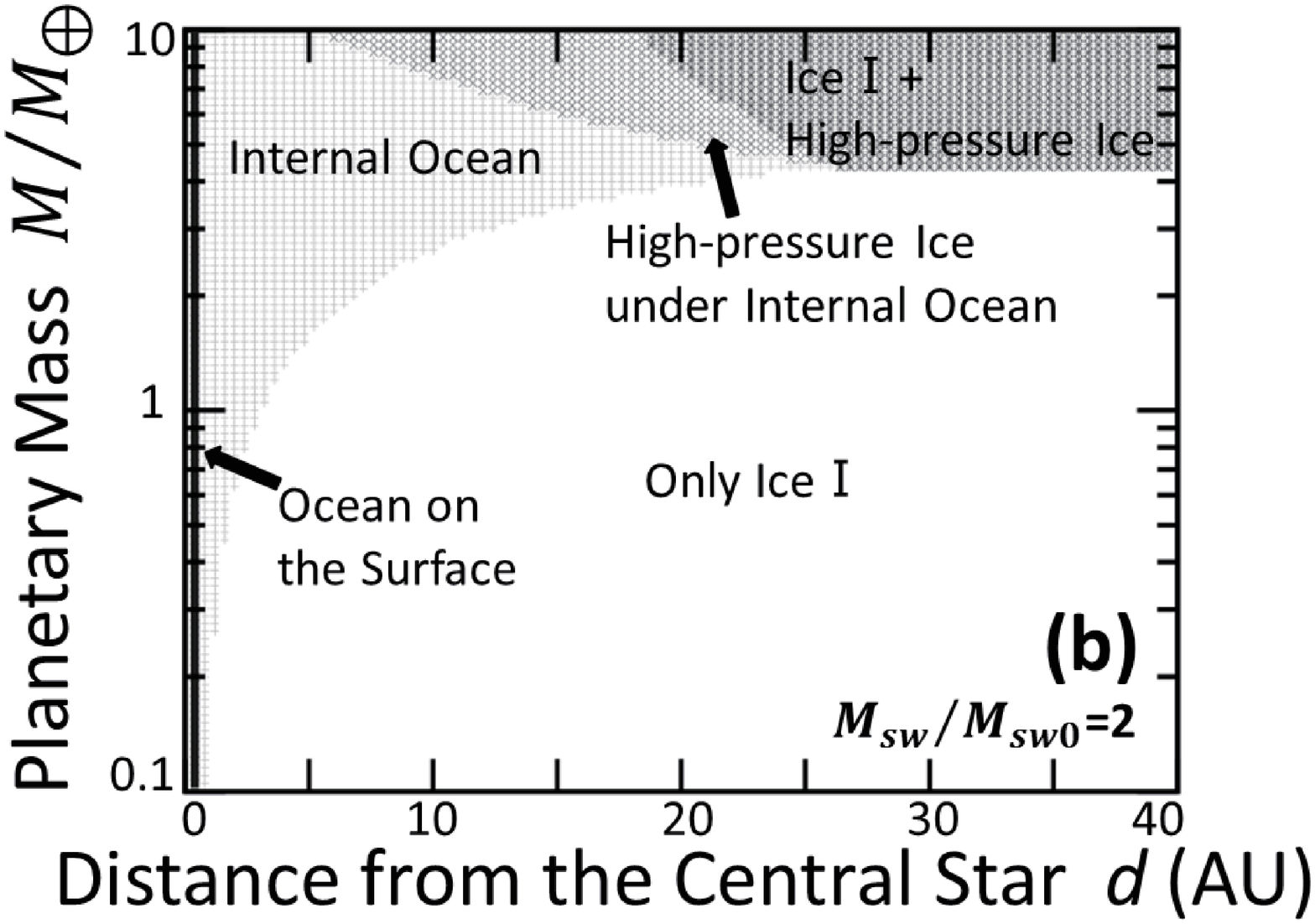}
\includegraphics[width=8cm,bb= 0 0 960 720,trim=0 0 0 180]{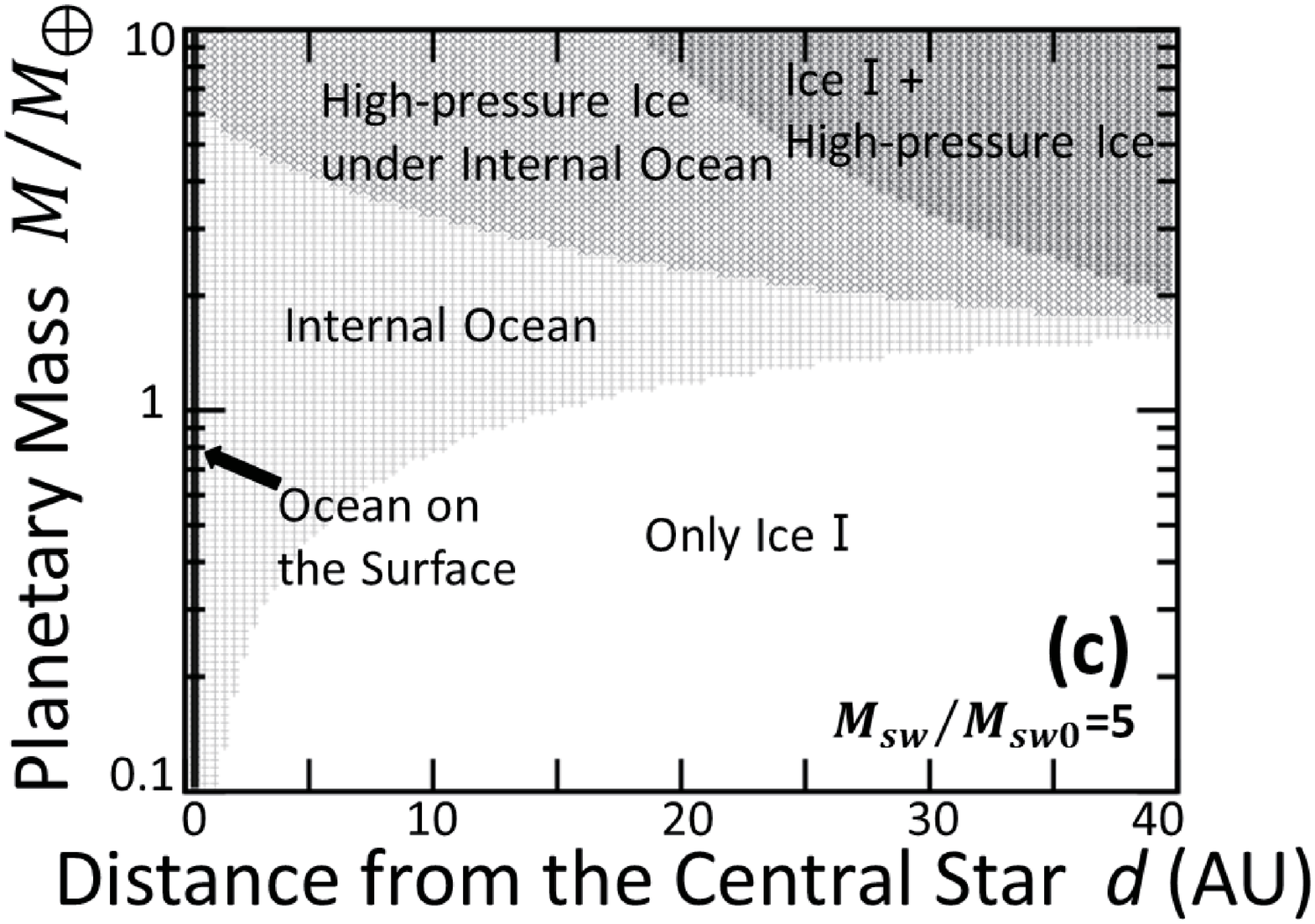}
\includegraphics[width=8cm,bb= 0 0 960 720,trim=0 0 0 180]{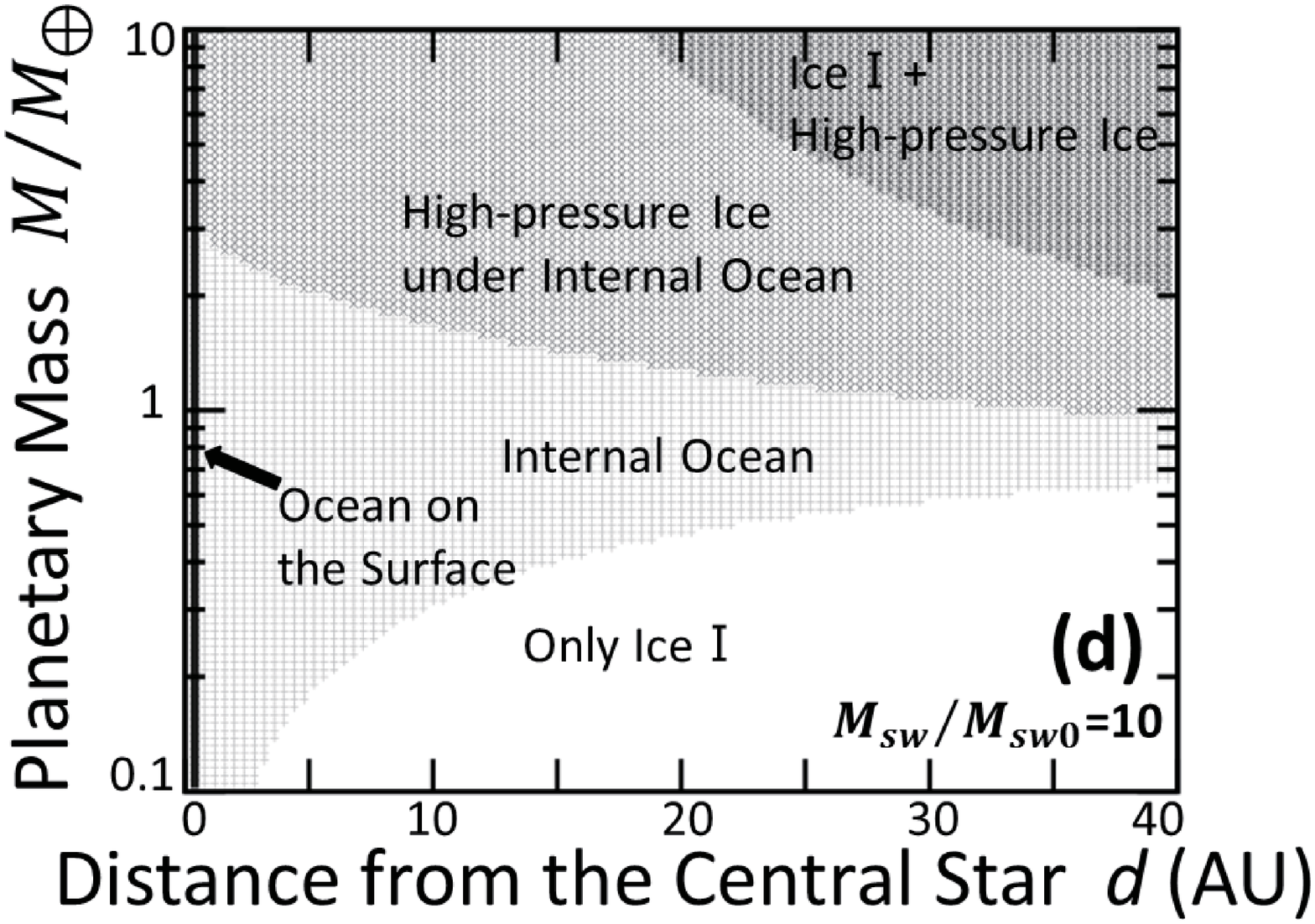}
\caption{Conditions for planet orbiting central star ($E/E_0 = 1$) for (a) $M_{sw}/M_{sw0} = 1$, (b) $M_{sw}/M_{sw0} = 2$, (c) $M_{sw}/M_{sw0} = 5$, and (d) $M_{sw}/M_{sw0} = 10$.  Horizontal axes represent the distance from the central star; vertical axes represent the planetary mass normalized by Earth's mass.}
\end{figure}

\clearpage

\begin{table*}
\caption[values]{Values of parameter for parameterized convection model}
\label{parameter}
\begin{center}
\begin{tabular}{llll}
\hline
Parameter & Value & Unit & Source\\
\hline
$R_m(1M_{\oplus})$ & $6,271 \times 10^3$ & m & \citet{blo08} \\
$R_c (1M_{\oplus})$ & $3,471 \times 10^3$ & m & \citet{blo08} \\
$\rho c$ & $4.2 \times 10^6$ & J m$^{-3}$K$^{-1}$ & \citet{blo08} \\
$T_m(0)$ & $3,000$ & K & \citet{blo08} \\
$k$ & $4.2$ & J s$^{-1}$m$^{-1}$K$^{-1}$ & \citet{blo08} \\
$R_{a_{crit}}$ & $1,100$ & ― & \citet{blo08} \\
$\beta$ & $0.3$ & ― & \citet{blo08} \\
$E_0$ & $1.46 \times 10^{-7}$ & J s$^{-1}$m$^{-3}$ & \citet{blo08} \\
$\lambda$ & $0.34$ & Gyr$^{-1}$ & \citet{blo08} \\
$g(1M_{\oplus})$ & $9.81$ & m s$^{-2}$ & \citet{blo08} \\
$\alpha$ & $3 \times 10^{-5}$ & K$^{-1}$ & \citet{blo08} \\
$\kappa$ & $10^{-6}$ & m$^2$s$^{-1}$ & \citet{blo08} \\
$\nu_m$ & $2.21 \times 10^{-7}$ & m$^{2}$s$^{-1}$ & \citet{mcg89} \\
$\alpha_1$ & $6.4 \times 10^4$ & K & \citet{fra95} \\
$\alpha_2$ & $-6.1 \times 10^{6}$ & K per weight fraction & \citet{fra95} \\
$M_m(1M_{\oplus})$ & $4.06 \times 10^{24}$ & kg & \citet{fra95} \\
$M_w(0)(1M_{\oplus})$ & $4.2 \times 10^{21}$ & kg & \citet{blo08} \\
$f_{bas}$ & $0.03$ & ― & \citet{blo08} \\
$\rho_{bas}$ & $2,950$ & kg m$^{-3}$ & \citet{blo08} \\
$d_{bas}$ & $5 \times 10^{3}$ & m & \citet{blo08} \\
$d_m$ & $40 \times 10^{3}$ & m & \citet{blo08} \\
$f_w$ & $0.194$ & ― & \citet{blo08} \\
$R_T$ & $29.8 \times 10^{-5}$ & K$^{-1}$ & \citet{blo08} \\
$ R_{H_2O,0}$ & $0.001$ & ― & \citet{blo08} \\
$A_0^*$ & $3.1 \times 10^{14}$ & m$^{2}$ & \citet{mcg89} \\
\hline
\end{tabular}
\end{center}
\end{table*}

\end{document}